\documentclass[11pt,preprint]{aastex}
\begin{document}

\title{LITHIUM IN TURN-OFF STARS” IN THE GLOBULAR CLUSTER M5: A QUEST
FOR PRIMORDIAL LITHIUM}

\author{Ann Merchant Boesgaard\altaffilmark{1}}
\affil{Institute for Astronomy, University of Hawai`i at M\-anoa, \\ 
2680 Woodlawn Drive, Honolulu, HI {\ \ }96822 \\ } 

\email{annmb@hawaii.edu}

\author{Constantine~P.~Deliyannis\altaffilmark{1}}
\affil{Department of Astronomy, Indiana University, 727 East 3rd Street, \\
Swain Hall West 319, Bloomington, IN {\ \ }47405-7105 \\ }

\email{cdeliyan@indiana.edu}

\altaffiltext{1}{Visiting Astronomer, W.~M.~Keck Observatory, jointly operated
by the California Institute for Technology and the University of California.}

\begin{abstract}
The light element lithium is formed by nucleosynthesis during the Big Bang.
Its abundance can help to define the parameters of the early universe.  To
find this primordial value, it is necessary to determine Li abundances in the
oldest stars because it is readily destroyed by nuclear reactions in stellar
interiors.  We have made high-resolution ($\sim$45,000) spectroscopic
observations of five identical unevolved turn-off stars in the 13 Gyr old
globular cluster M5.  In our analysis we find a range in Li abundance of a
factor of two; the spread is five times the individual error.  The comparison
of these results with those for turn-off stars from five other globular
clusters reveals similarly large range in Li.  Lithium in M5 and the other
clusters all have stars above the field star Li plateau, but none are as high
as the predictions for primordial Li.  The maximum values for Li are the same
in all six clusters.  Multiple generations of stars are found in many
globular clusters; those later generations are expected to have formed from
Li-depleted gas.   Such second- and later-generation stars would have no
Li.  However, only one of the six clusters has a few unevolved stars
with upper limits on the Li abundance.
\end{abstract}

\section{INTRODUCTION}
The study of the element Li has provided information on nucleosynthesis during
the Big Bang.  Along with $^2$H, $^3$He, and $^4$He, $^7$Li is produced during
the first 15 minutes of the Universe by neutrons and protons.  However, Li is
easily destroyed in stars and can also be created in the interstellar medium
through spallation by energetic (150 MeV) particle collisions with abundant C,
N, O atoms.  To establish the value of the primordial Li it is necessary to to
measure it in the oldest, unevolved stars in the Galaxy.  The earliest attempt
at this was the study of Li in the most metal-poor field stars by Spite \&
Spite (1982).  A recent example extends Li abundances down to a star having
[Fe/H] = $-$6.1 (Aguado et al.~2019).  Other very metal-poor stars in the
galactic halo, both dwarfs and giants,  have been studied for Li and other
elements by Bandyopadhyay et al.~(2022).

The unevolved stars in the oldest globular clusters provide another venue for
establishing the primordial value of Li.  One of the oldest and most
metal-poor globular clusters is M 92.  In a compilation of 68 globular
clusters by age Valcin et al.~(2020) give 13.30 $\pm$0.06 Gyr for M 92.
Recently, Ying et al.~(2023) determine the absolute age of M92 to be 13.80
$\pm$0.75.  Bailin (2019) tabulates the metallicities for 55 globular
clusters and gives [Fe/H] = $-$2.239 $\pm$0.028 for M 92.  We reported on the
Li abundances in several turn-off stars in M 92 (Deliyannis et al.~1995,
Boesgaard et al.~1998).  We found a range in Li abundance of a factor of three
in our five stars with similar temperatures, all near 5800-5900 K.  The spread
in A(Li) = log N(Li) - log N(H) +12.00 was 2.01 to 2.45 which was attributed
to differences in initial angular momenta and then spin-down.  The slowest
rotators would lose the least Li while the rapid rotators would undergo more
Li depletion via mixing to hotter interior regions where Li would be destroyed
by reactions with protons.

Large studies of Li in globular clusters that include turn-off stars have been
done for four southern globular clusters.  Aoki, Primas, Pasquini
et al.~(2021) studied 357 stars from the turn-off to red giants in 47 Tuc. Lind et al.~(2009) observed Li in 349 stars in NGC 6397 of which 180 were
turn-offs stars, while Gonzalez Hernandez et al.~(2009) found Li abundances
in 163 stars in that cluster.  NGC 6752 was studied for Li by Shen, Bonifacio,
Pasquini et al.~(2010) in 112 turn-off stars.  Both Monaco et al.~(2012)
and Spite et al.~(2016) examined Li abundances in 91 main-sequence and
subgiant stars in M4.

Unevolved stars in globular clusters contain information about the basic
composition of the stars before they have undergone any mixing with their
internal layers during post-main-sequence evolution.  One important caveat to
this idea is that multiple generations of star formation have occurred as
discussed in a review by Gratton et al.~(2012).  Their Figure 6 shows examples
of the well-known Na-O anti-correlation in 20 globular clusters.  Early
generations of massive stars that explode into supernovae in star-forming
regions will enrich that next generations.  This could result in a spread in
metallicity even in the first population of stars.  Lardo, Salaris, Cassini et
al.~(2023) found evidence for this in NGC 2808.  

Here we are studying Li in the main-sequence turn-off stars in M5 for the
first time with high resolution spectra.  These stars have V = 18, some 2-6
magnitudes fainter than the red giants.  There is excellent CCD photometry of
M5 by Sandquist et al.~(1996).  They find an age for M5 of 13.5 $\pm$1 Gyr
with [Fe/H] = $-$1.40 (Zinn \& West (1984) and 11 $\pm$1 Gyr with [Fe/H] =
$-$1.17 (Sneden et al.~(1992).  The most quoted value of [Fe/H] is $-$1.33
from Carretta et al.~(2009), and is between those two.  Although M5 is an old
globular cluster (12.75 $\pm$0.08 Gyr Valcin et al.~2020), it has a
metallicity at the median for the 19 clusters compiled by Carretta et
al.~(2009) and is given a value for [Fe/H] = $-$1.259 $\pm$0.003 in the
collection of Bailin (2019).

A special feature of this research is that the five stars are virtually
identical in V, (B-V)$_0$, T$_{\rm eff}$, log g.

Lithium abundances in M5 have been reported for 99 red giant stars by D'Orazi
et al.~(2014) with spectral resolution of 17,000.  They find values for A(Li)
predominantly between 0.8 and 1.1.  This represents the dilution of Li in red
giants caused by the expansion of the surface convection zone.

\section{OBSERVATIONS AND STELLAR PARAMETERS}

The turn-off stars in M5 have V = 18.0 but it was possible to observe them at
high spectral resolution with HIRES (Vogt et al.~(1994) on the Keck I
telescope.  Figure 1 shows the fit through the data for the color-magnitude
diagram for M5 from Sandquist et al.~(1996).  The positions of the nearly
identical five stars we observed are shown by red stars at V = 18.0.

Multiple exposures were required to obtain high enough S/N ratios, but with a
limit of 45-60 minutes on each exposure.  The exposure times had to be long
enough to obtain good signal, but not so long that they were subjected to too
many cosmic-ray events.  On each night there were at least two exposures of
the Th-Ar lamp, one at the beginning of the night and one at the end.  We took
15-20 quartz flat-field frames and 15-20 bias frames.

The log of our spectral observations of five turn-off stars in M5 appears in
Table 1.  The photometry is from Sandquist et al.~(1996).  The total observing
time for each star of the co-added spectra is between 195 min. and 345 min.
The echelle spectra have 29 orders and cover from 4425 to 6880 \AA{} with some
inter-order gaps at longer wavelengths.  .  The spectral resolution is
$\sim$45,000 or $\sim$0.04 \AA{} pix$^{-1}$ in the Li I region.  The data
reduction was done with IRAF(Tody 1986, 1994)$\footnote{IRAF is distributed by
the National Optical Astronomy Observatories, which are operated by The
Association of Universities for Research in Astronomy, Inc. now NOIRLab, under
cooperative agreement with the National Science Foundation.}$ including
bias-subtraction, flattening, wavelength calibration from the Th-Ar reference
lamp.  The individual spectra for each star were wavelength adjusted, then
co-added.  Those co-added spectra were then fit with a continuum with IRAF.

We show the Li region of the co-added, continuum-fit wavelength corrected
spectra for two of the stars in Figure 2.  These spectra have comparable S/N
ratios.  The temperatures of the two stars agree with each other within 25 K.
The positions of the Li I and Ca I lines are indicated.  While the Ca I line
strengths are seen to be similar in the two stars, the Li I line is clearly
stronger in 5323 than in 5351.

We have measured equivalent widths of several Fe I lines in those same two
stars, 5323 and 5351, and show those results in Figure 3.  The 45 degree line
shows where the measurements would be equal for the two stars.  The median
difference is 5.4 m\AA{} for each pair of lines.  The difference for the Li I
line in the two stars is 19.9 m\AA.

In our past work on Li abundances in globular clusters we have used
temperatures scales derived by Carney (1983) and King (1993) from (B-V)
colors.  We now include a temperature scale from Ramirez \& Melendez (2005)
based on the infrared flux method (IRFM).  The King scale results in the
hottest temperatures and the Ramirez \& Melendez the coolest with a difference
of $\sim$210 K.  Each of our five stars are close to each other in temperature
on any of the scales.  For a given temperature scale our five stars differ by
less than 60 K and less than 0.05 in log g.  We can conclude that they are
quintuplets.
  
\section{LITHIUM ABUNDANCES}

We have determined Li abundances with the spectrum analysis program
MOOG\footnote{http://www.as.utexas.edu/$~$chris/moog.html} (Sneden 1973,
Sneden et al.~2012).  We used the stellar parameters in Table 2 on all three
temperature scales.  Abundances of Li were found by spectrum synthesis with
${\it synth}$.  

Examples of the synthesis fits for two stars are shown in Figure 4 from the
Carney (1983) temperature scale.  The best fit syntheses show that these two
stars differ in Li by a factor of 1.8.  Our line list includes the Li doublet
and hyperfine structure.  It also includes a line of Fe I at 6707.441 which
contributes virtually nothing at these temperatures.  We also found the Li
abundances with the ${\it blends}$ feature in MOOG from the measured
equivalents widths.

The results are given in Table 3 for all five stars with all three temperature
scales for both the synthetic spectrum results and the equivalent with
measurements.  We prefer the results from the spectrum synthesis, but note the
agreement between the two techniques is very good with a mean difference of
0.05 in A(Li).

The Li abundances are sensitive to temperature, but not to the other model
parameters.  A change of +60 K results in a change in A(Li) of +0.04 dex.  The
two stars shown in Figure 4 differ in T$_{\rm eff}$ by 58 K and in A(Li) by
0.25 dex or six times greater than errors due to temperature.

it is of interest to examine further whether the apparent differences in A(Li)
between these stars are real, as opposed to simply reflecting errors.  We
proceed in a manner similar to that in our M92 work.

The stars were deliberately chosen to have near-identical temperatures, so
that differences in A(Li) would simply be a reflection of differences in the
equivalent width of the Li line.  We then apply the Cayrel (1988) formulation
for calculating errors in equivalent widths, as recast by Deliyannis et
al.~(1993).  The 1-sigma error due to photon noise alone, $\sigma$$_{ph}$,
depends on the width of the line and the S/N ratio (SNR) per pixel.  We assume
that the 1-sigma error due to continuum placement, $\sigma$$_{co}$, is of
similar magnitude and then add that in quadrature to $\sigma$$_{ph}$ to get
the total a total 1-sigma error, $\sigma$$_{tot}$.  As a check of these
procedures and assumptions, we measured 14 Fe I lines on the linear part of
the curve of growth in stars 5323 and 5351 (see Figure 3) and found an
average difference of 6.54 mA, or 4.62 m\AA{} per measure.  The average width
of these lines is 137 m\AA{} (with a dispersion of 47 m\AA{} per pixel), and
the SNR per pixel is nearly identical for both stars, 37 and 35, respectively.
This yields $\sigma$$_{ph}$ = 3.35 m\AA{}, and $\sigma$$_{tot}$ = 4.73 m\AA{},
nearly identical to the value 4.62 m\AA{} indicated above.

The width of the Li I doublet is 185 m\AA{}, a bit larger than that for the Fe
I lines, as expected, and thus yields slightly larger errors for these stars
of $\sigma$$_{ph}$ = 3.89 m\AA{}, and $\sigma$$_{tot}$ = 5.50 m\AA{}.  The
measured Li equivalent widths are 53.7 and 33.8 m\AA{}, respectively, which
are solid detections at the 9.8 $\sigma$ and 6.1 $\sigma$ level, respectively;
their difference is 19.9 m\AA{}.  A chi-squared test suggests that there is
only a 0.00030 probability of obtaining the measured equivalent widths purely
by chance with the given errors (and assumptions) if the real equivalent
widths are identical.

We repeat this analysis for stars 5287 and 5364, which have an even smaller
difference in temperature (about 8 K) than stars 5323 and 5351 (about 23 K).
But the difference in SNR per pixel is larger: 39 and 32, respectively.  Once
again, the measured Li equivalent widths are strong detections at the 10.5 and
5.3 $\sigma$ level, respectively.  For simplicity, we adopt SNR = 32 for both
stars, which may overestimate the error for star 5287.  Then, a chi-squared
test suggests that there is only a 0.00094 (or smaller) probability of
obtaining the measured equivalent widths purely by chance with the given
errors if the real equivalent widths are identical.

Whether we compare star 5323 to 5351, or star 5287 to 5364, there is strong
evidence that there is a real dispersion in A(Li) in our M5 sample.

For our M5 stars the corrections for effects of NLTE and 3D on Li line
formation are small.  Wang et al.~(2021) calculated these effects on three Li
I lines including the resonance line, $\lambda$6707 of our observations.
Their results, as highlighted in their Figure 8, indicate corrections of less
than $-$0.04 in A(Li)  for our range in T$_{\rm eff}$, log g and [Fe/H].

\section{DISCUSSION}

Both Figure 4 and Table 3 show that there is a range in Li abundances in our
five otherwise-identical stars.  The spread is a factor of two.  All but one
of the stars has a Li amount greater than the field star plateau of A(Li) =
2.2. These results are similar to our findings for Li in six turn-off stars
in M 92 (Boesgaard et al.~1998) with a range in A(Li) of 0.44 dex and half of
them above the field star plateau.

Abundances of Li have been determined in other globular clusters.  NGC 6397
was studied for Li first by Pasquini \& Molaro (1996) in six stars with three
near the turn-off.  Later Bonifacio, Pasquini, Spite et al.~(2002) determined
Li abundances of 12 turn-off stars.  There were 79 main-sequence stars
included in the research by Gonzalez Hernandez, Bonifacio, Caffau et
al.~(2009).  In addition, Koch, Lind \& Rich (2011) reported a turn-off star
that is super Li rich with A(Li) = 4.03.

For NGC 6752 there were nine turn-off stars in the Li study by Pasquini,
Bonifacio, Molaro et al.~(2005).  Schiappacasse-Ulloa, Lucatello, Rain et
al.~(2022) found Li abundances in 217 stars of which 156 were turn-off and
subgiant branch stars.  Their range for Li in the turn-off stars was A(Li) =
2.4 down to an upper limit of 1.6.   Gruyters et al.~(2014) looked into the
effects of atomic diffusion on Li (and several other elements) in 194 stars
from the turn-off to red giants.  For the turn-offs they found A(Li) values
from 2.1 to 2.4 with three stars with only upper limits on Li.  Shen et
al.~(2010) found a correlation between Li and O in a sample of 112 turn-off
stars and suggest that there would be Li production in the polluting gas.

The abundance of Li has been determined in 91 stars in the globular
cluster M4 by Monaco et al.~(2012) and Spite et al.~(2016).  Of those stars
there are 52 with temperatures above 5800 K and log g larger than 4.00.  They
show a range in A(Li) from 1.82 to 2.40.  They comment on one star in this
range that has a ``remarkably'' large Li value at 2.87 and discuss
possibilities for this at some length.

One of the most metal-rich globular clusters is 47 Tuc = NGC 104.  Pasquini \&
Molaro (1997) found Li abundances in two turn-off stars and Bonifacio et
al.~(2007) determined Li with high-resolution spectra of four turn-off stars.
A major work on Li was done by Aoki, Primas, Pasquini et al.~(2021).  They
found Li abundances for 93 turn-off stars in their total sample of 347 stars.
Their range in A(Li) was 1.5 - 2.3 for the turn-off stars.

A compilation of the ages of globular cluster has been done by Valcin et
al.~(2020) and of [Fe/H] vales by Bailin (2019).  We show these values for the
six clusters in Table 4.  (Note that the Planck collaboration age of the
Universe is 13.7 Gyr while the age given by Valcin et al. is 14.21 Gyr.)  The
results for A(Li) in turn-off stars are shown in Figure 5 as a function of the
cluster Fe abundance.  Of the six clusters M5 is the youngest at 12.75
Gyr, but cluster age does not seem to affect the measured Li abundance or the
spread in Li in turn-off stars.

We see these indications from Figure 5:

1) All six clusters show a large range in A(Li) in these unevolved
   stars.

2) All have many stars above the field star Li plateau.

3) All of these unevolved stars have Li detections; except for a few stars in
NGC 6752, none have upper limit values for A(Li).

4) None has an amount of Li that is near the predictions for primordial Li.

5) The maximum A(Li) abundance for all six clusters is similar, near 2.5
dex.

The explanation for the values and the range in A(Li) values that we proposed
for M 92 (Boesgaard et al.~1998) was that differences in initial angular
momentum played an important role in the preservation, or lack thereof, of the
surface content of Li.  The most rapid rotators would spin down the most and
destroy the most Li.  The slower rotators would not spin down as much and so
preserve more Li.  All the stars may have lost some of their initial Li from a
potentially higher initial amount.

This ``rotational mixing'' could account for the spread in the Li content.
The two stars with the lowest A(Li) (stars 5351 and 5364)  have lower A(Li)
($\sim$2.22) than the two stars 5287 and 5303) with the highest A(Li)
($\sim$2.43).  It could also account for the result that the Li abundances of
all our stars lie below the value A(Li) = 2.7-2.8 inferred from Planck data in
the context of standard Big Bang Nucleosynthesis.

The formalism of how models with a variety of rotational instabilities should
be constructed was pioneered by Endal \& Sofia (1976, 1981).  The the
Yale-style models predict that such depletion may vary slightly from star to
star because of differences in the initial stellar angular momenta
(e.g.~Pinsonneault et al.~1989, 1992, Somers \& Pinsonneault 2016).

This rotational mixing has explained many observations in Populations I
dwarfs about the surface depletion Li, Be, and B.  For example: Higher than
normal Li abundances were observed in short period tidally locked binaries
(Thorburn et al. 1993, Ryan \& Deliyannis 1995); the cool side of the ``Li
Dip'' discovered by Boesgaard \& Tripicco (1986); the correlated depletion of
Li and Be (Deliyannis et al. 1998; Boesgaard et al. 2004) and of Be and B in F
dwarfs (Boesgaard et al. 2005, 2016); the Li-temperation relation and the
Li/Be ratio of M67 subgiants evolving out of the cool side of the Li Dip
(Sills \& Deliyannis 2000; Boesgaard et al. 2020); the correlation between
stellar spindown and Li depletion in late A/early F dwarfs (Deliyannis et
al. 2019); and the continued depletion of Li during the main sequence in G
dwarfs (Boesgaard et al. 2022, Sun et al. 2023).  Explanation of Li
abundances in late G/K dwarfs may in addition require magnetically-induced
radius inflation (Somers \& Pinsonneault 2015; Jeffries et al. 2021).
 
Conclusion 3) above is more difficult to explain.  When high-mass,
first-generation stars evolve and become supernovae, they destroy their Li.
So the newly formed stars would be without Li.  Those second-generation stars
-- main-sequence and turn-off stars -- would not have detectable Li.  The
expectation is that there would be many stars with only upper limits on A(Li).
Instead, only in NGC 6752 do a few turn-off stars of the 217 have only upper
limits.  This issue is highlighted by Shen et al.~(2010) in their work about
the Li-O anti-correlation in NGC 6752.  The polluting gas would have to be
enriched in Li.  The lack of lithium-less stars among main-sequence and
turn-off stars in these globular clusters is one of many unexplained
observational puzzles regarding multiple populations in globular clusters.
Several anomalies have been discussed in the review by Bastian \& Lardo
(2018).

Although all of these clusters have similar maximal values of A(Li), that
value is lower than the predictions from Big Bang nucleosynthesis.  This issue
has been discussed by Deal \& Martins (2021).  That maximum value, about 2.5,
could represent a common amount of depletion due rotational effects, as
mentioned above.  It could also actually be the primordial amount and the flaw
is in some aspect(s) of the predictions.

\section{SUMMARY AND CONCLUSIONS}
 
One method to assess the primordial amount of $^7$Li is to determine its
abundance in the oldest, most unevolved stars.  Inasmuch as Li is destroyed in
the interiors of stars and diluted at the surfaces of red giants, Li
abundances in those evolved stars is not representative of the initial
content.  The turn-off stars in old and metal-poor globular clusters are
plausible subjects for such an investigation.  (Main-sequence stars in
globular clusters are usually too faint for quality spectral analysis.)  In
this work we have obtained and analyzed high-resolution spectra of five such
stars in the globular cluster M5.  These stars are very faint at V = 18.0.
With the Keck I telescope and HIRES we obtained and co-added multiple
exposures of each of the five stars.  The spectral resolution we obtained was
$\sim$45,000 or 0.04 \AA{} pix$^{-1}$.

We selected stars to be very closely similar to each other.  The temperatures
of our five stars agree to within 60 K.  In spite of that, we found their Li
contents range over a factor of two with that spread being five times the
observational error.  

We compared these results for M5 with our earlier findings for Li in six
turn-off stars in M92 (Boesgaard et al.~1998).  Those similar stars showed an
even larger array of Li values.  The comparisons with results for six
clusters were compiled as a function of their metallicities, [Fe/H], in Figure
5.  All show a wide range in A(Li), but all have a maximum near A(Li) = 2.5.
This is higher than the field star plateau of $\sim$2.2, but lower than the
predicted amount for primordial Li of 2.7-2.8.

The interpretation for the spread in Li values is that the stars had different
initial angular momenta and as they spun down, they circulated surface Li down
to higher temperatures where it would be destroyed by nuclear reactions with
protons, forming 2 He nuclei.  The faster rotators would circulate and destroy
more Li.  It is plausible that all of the stars destroyed some Li from a
higher initial amount than the measured 2.5.

If any of the turn-off stars were second or third generation cluster members,
they would be expected to have no Li.  The progenitors would have diluted or
destroyed their initial Li.  Instead, there are only a few unevolved stars in
only one of the six clusters, NGC 6752, with upper limit values for A(Li).

Unevolved stars in these old, very metal-poor globular clusters uniformly show
greater amounts of Li than found in the oldest, metal-poor field stars.  For
each cluster there is a range in Li abundances of at least a factor of two.
The maximum Li amount is the same in all of them at A(Li) $\sim$2.5.  This
amount, however, is lower than the predictions from Big Bang nucleosynthesis
of 2.7-2.8.  It is of interest to extend this study to additional globular clusters.

\acknowledgments We wish to express our appreciation the Keck Observatory
support astronomers for their assistance and knowledge during our observing
runs.  CPD is grateful for the support through the National Science Foundation
grant AST-1909456.

Facility: Keck I HIRES Software; IRAF (Tody 1986, 1993);MOOG (Sneden 1973;
Sneden et al. 2012)

ORCID IDs Ann Merchant Boesgaard https://orcid.org/0000-0002-8468-9532
Constantine P. Deliyannis https://orcid.org/0000-0002-3854-050X

\begin{figure}
\plottwo{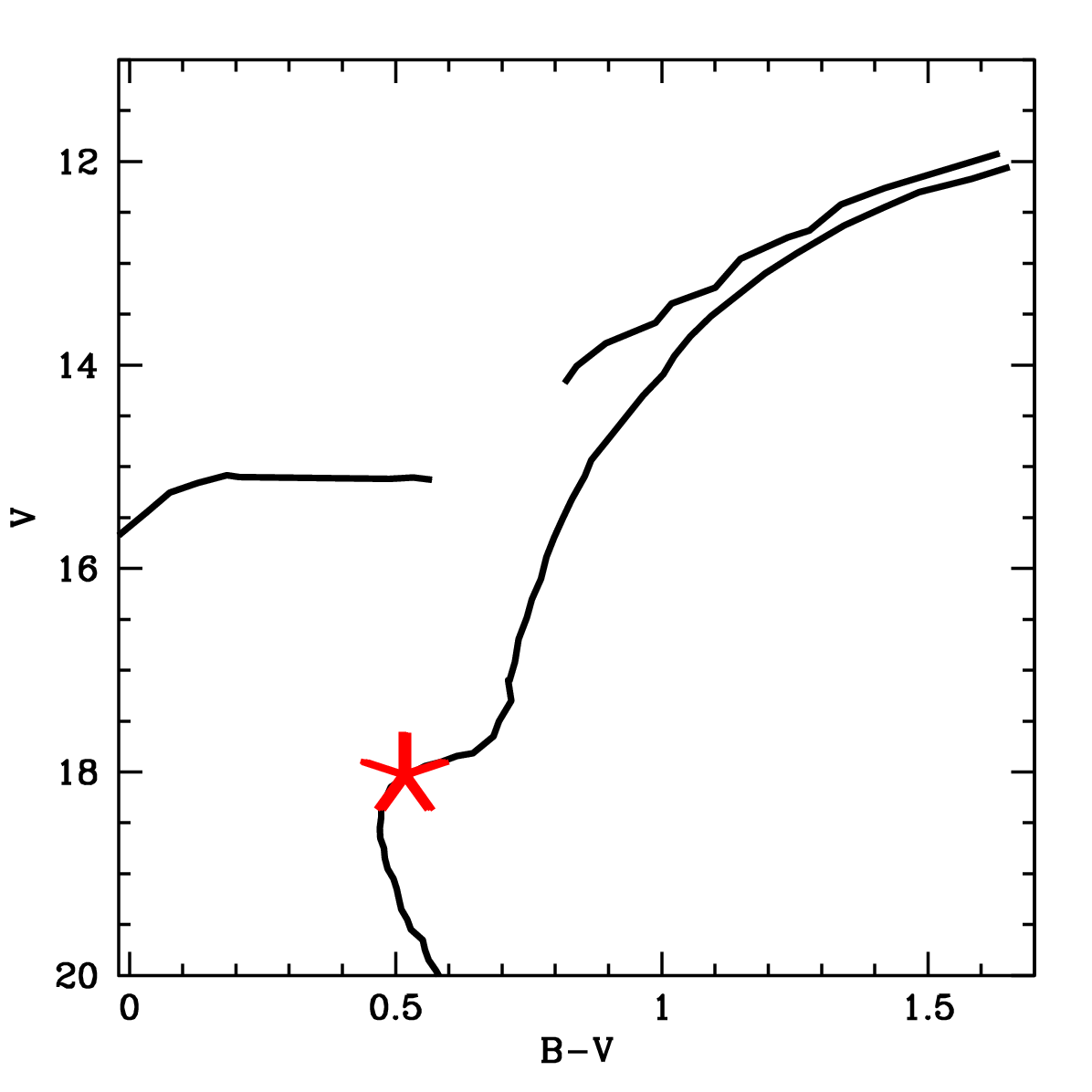}{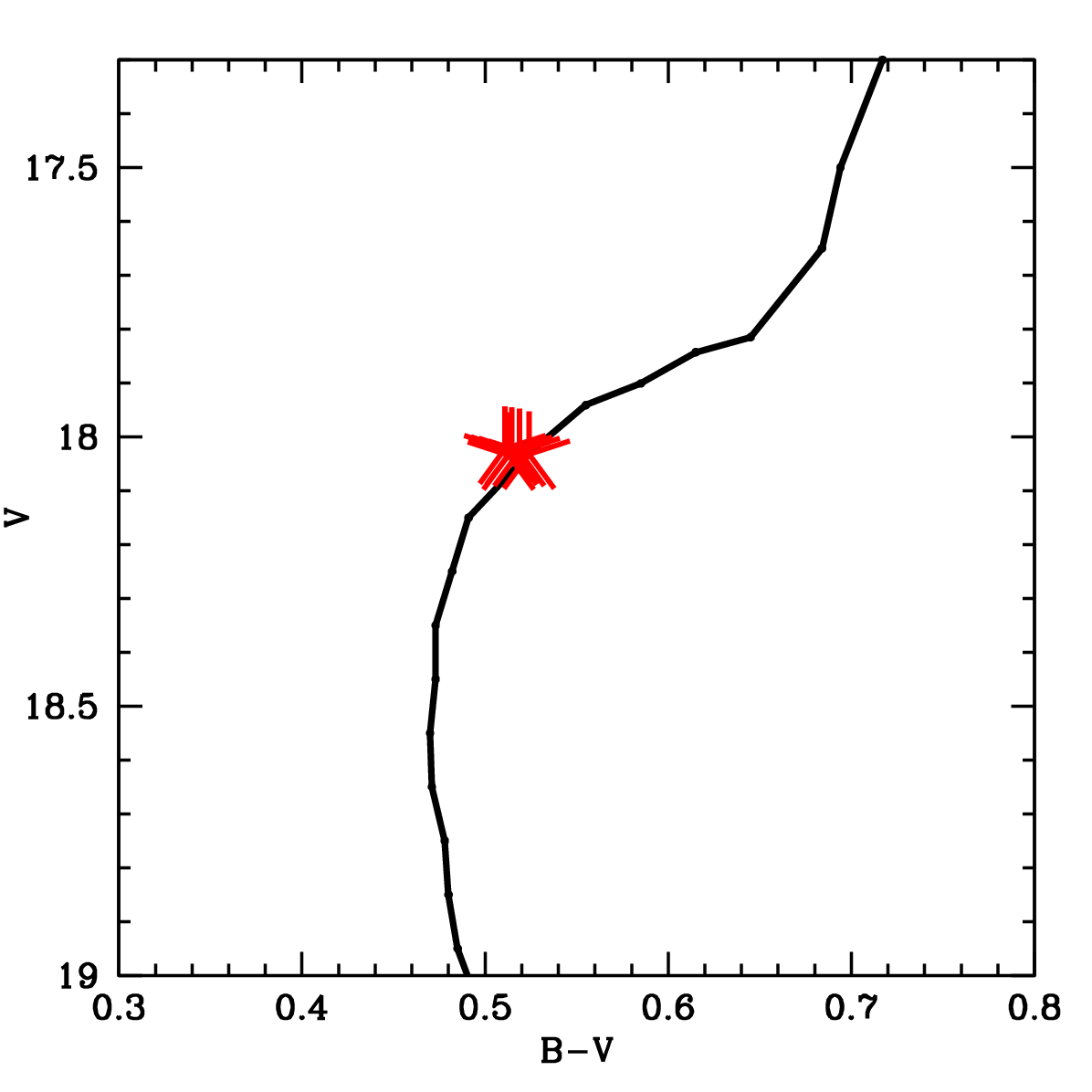}
\caption{Color-magnitude diagram for M5 from Sandquist et al.~(1996).  Left:
The full diagram showing their best fit to the observations.  The stars we
observed are shown as red stars.  Right: An enlargement of the main sequence
and turn-off sections of their C-M diagram with our stars as red stars.  Note
the faintness of our five stars and their similarity.}
\end{figure}

\begin{figure}
\plotone{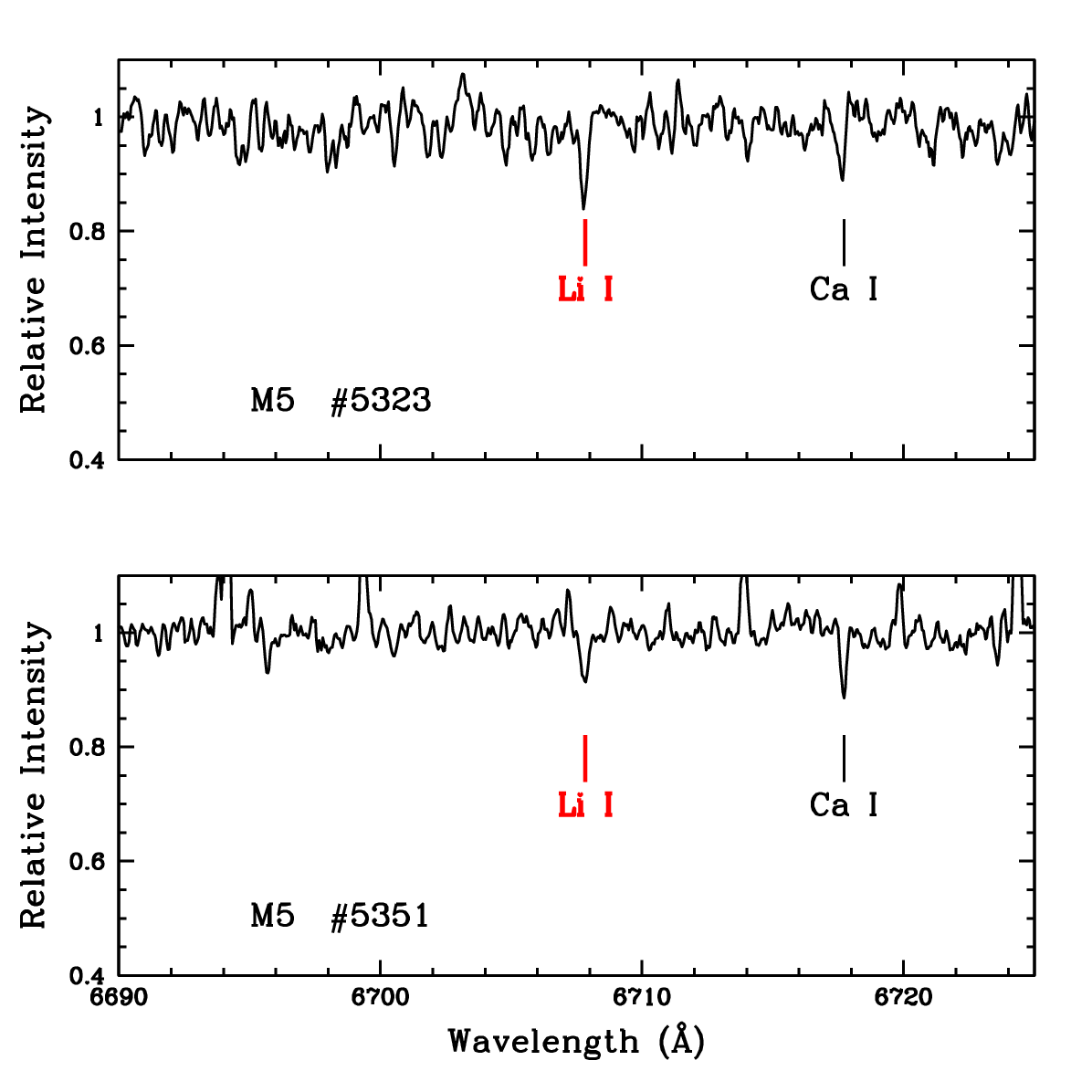}
\caption{Examples of spectra of two stars in the Li region.  The positions of
the Li I and Ca I features are indicated.  The Ca I lines are similar in
strength in both stars, while the the Li I feature is stronger in 5323.}
\end{figure}

\begin{figure}
\plotone{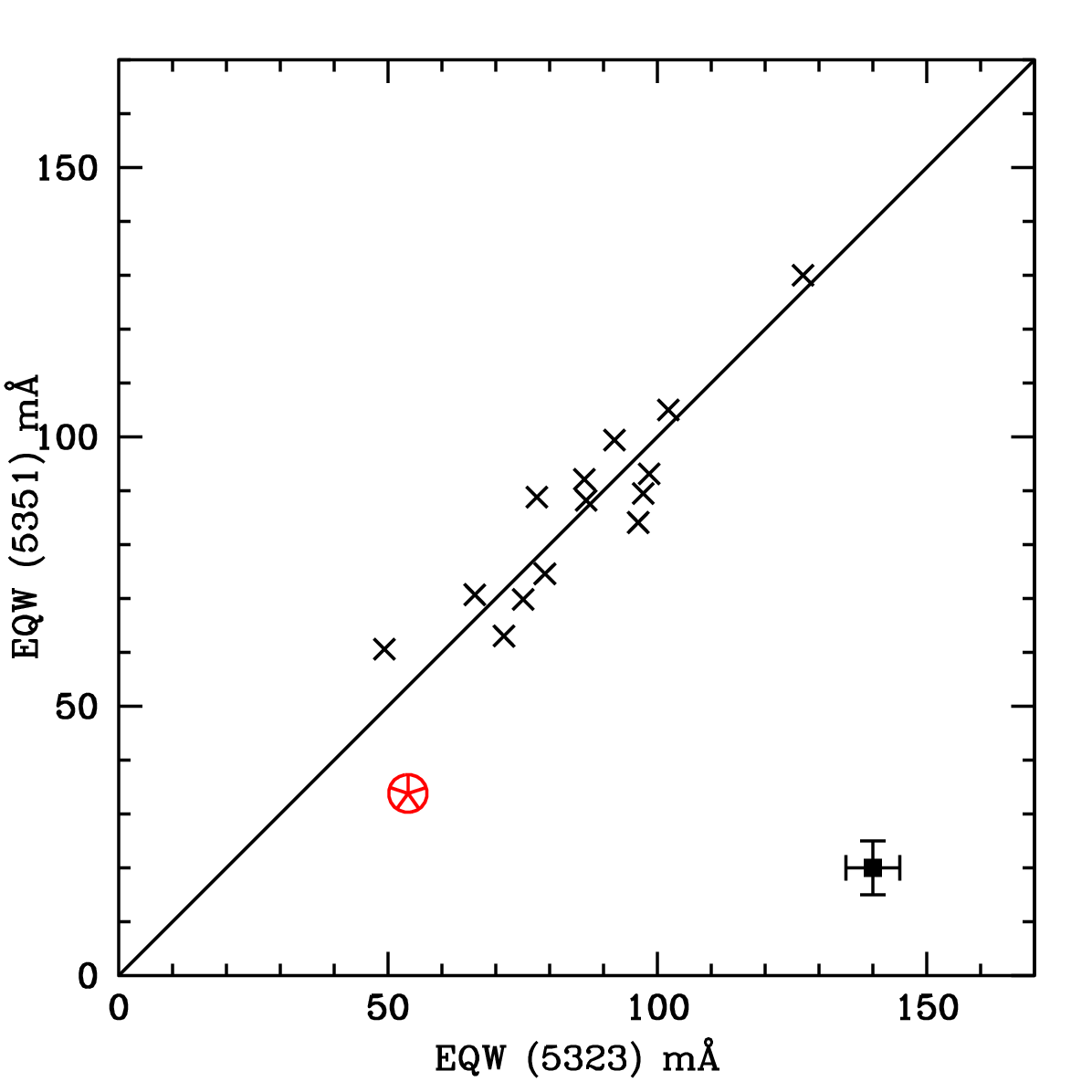}
\caption{The equivalent width measurements for two stars in M5.  The crosses
are for several Fe I lines and the red-circled star is for the Li I line which
is stronger in the star 5323.  The line indicates equality in the equivalent
widths.  An error bar for the line strength measurements is shown in the lower
right and corresponds to $\pm$5 m\AA.}
\end{figure}

\begin{figure}
\plotone{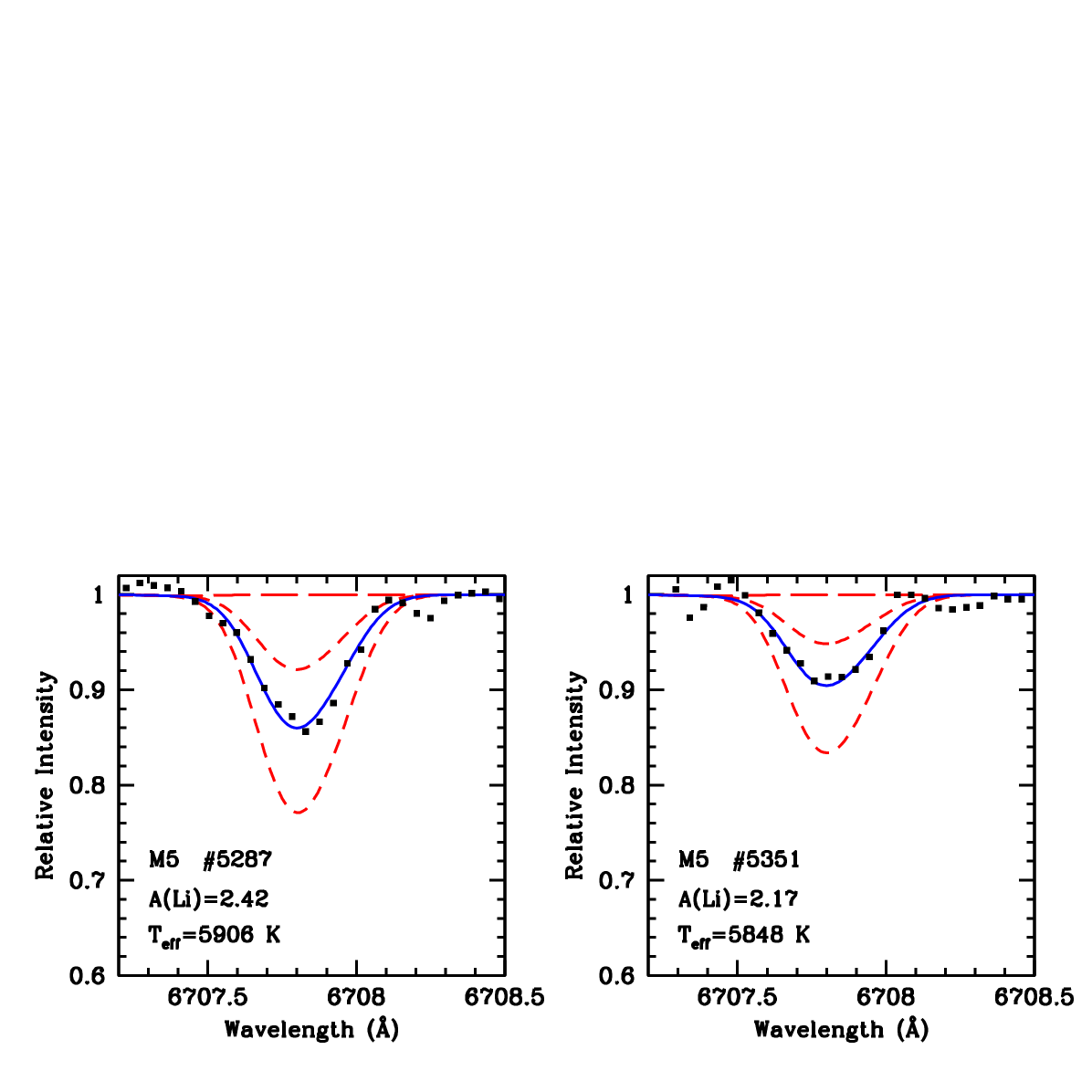}
\caption{The spectrum synthesis fits for Li in two of our stars.  The black
squares are the observed spectra and the blue solid line is the best fit.  The
red dashed lines are a factor of two more Li and a factor of two less Li; the
long-dashed red line represents no Li at all.  The temperatures and A(Li)
values are for the Carney (1983) temperature scale.}
\end{figure}

\begin{figure}
\epsscale{0.7}
\plotone{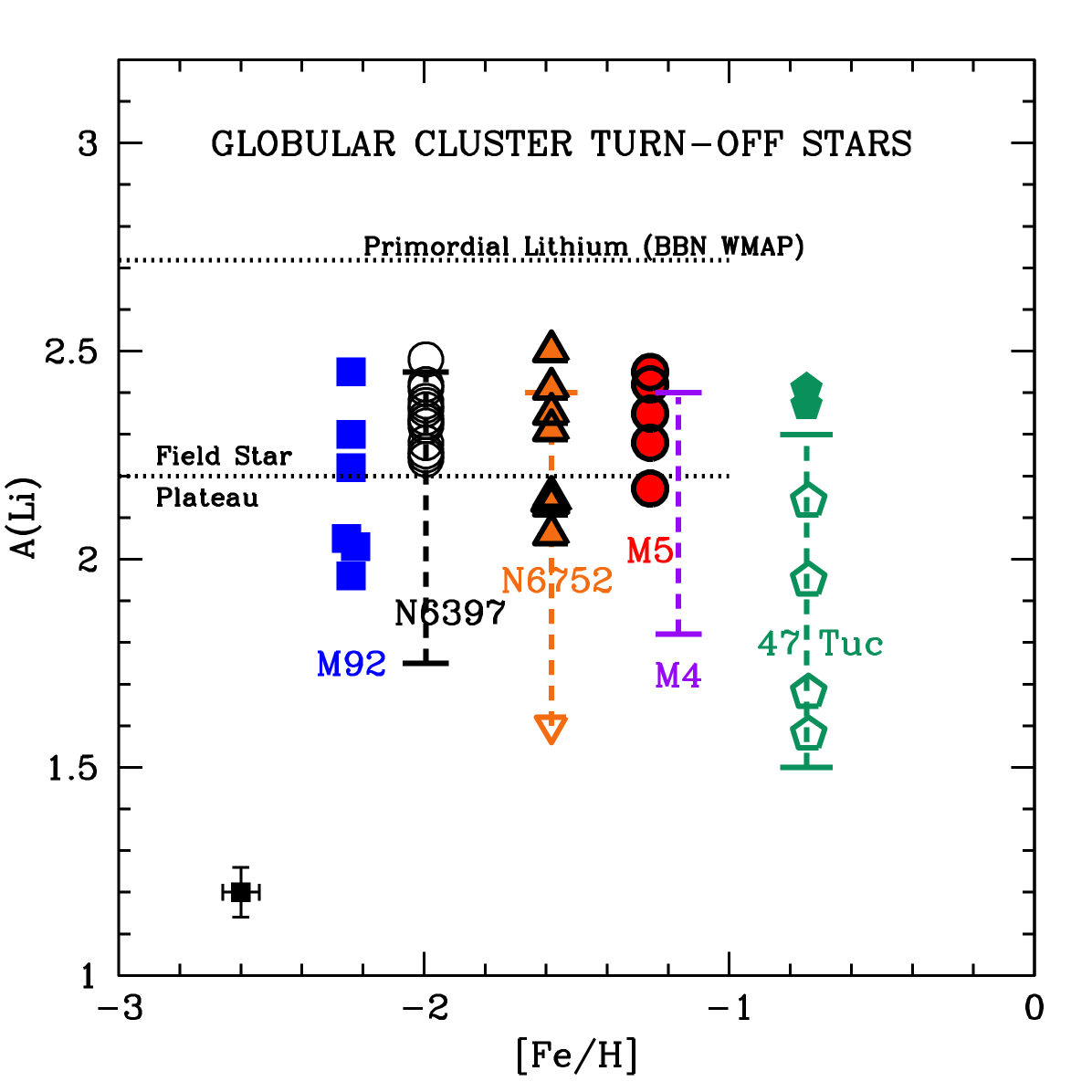}
\caption{Lithium abundances in turn-off stars in several globular clusters
with a range of iron abundances.  The results for M5 are from this work on the
Carney temperature scale shown as filled red circles.  The results for M92 are
from Boesgaard et al.~(1998) also on the Carney temperature scale are shown as
blue filled squares; two stars with identical Li values are offset from each
other for clarity.  The values for NGC 6397 are from Bonifacio et
al.~(2002)and Pasquini \& Molaro (1996) shown as open black circles.  The
range in values for NGC 6397, shown by the vertical black dashed line, are
from Lind et al.~(2009) and Gonzales Herndez et al.~(2009).  The orange
triangles are for NGC 6752 represent nine stars from Pasquini et al.~(2005).
The range for 217 turn-off stars, shown as the vertical dashed line is from
Gruyters et al.(2014) and Schiappacasse-Ulloa et al.~(2021).  There are 52
turn-off stars in M 4 shown as a range in A(Li) as the purple dashed vertical
line from Monaco et al.~(2012) and Spite et al.~(2016).  The values for 47 Tuc
(NGC 104) are from Pasquini \& Molaro (1997) shown as green filled pentagons
and those by Bonifacio at al (2007) by open green pentagons.  The green dashed
line shows the range of values for turn-off stars for 47 Tuc found by Aoki et
al.~(2021).  A typical error bar is shown in the lower left corner.  The
horizontal line at A(Li) = 2.2 represents the mean abundance of Li found for
metal-poor field stars first identified by Spite \& Spite (1982).  The
horizontal line at A(Li) = 2.719 corresponds to the predicted value for
primordial Li indicated by the WMAP results (Cyburt et al.~2008 and Cyburt et
al.~(2016).}
\end{figure}

\clearpage

%\documentclass[12pt,preprint]{aastex}
%\begin{document}
\begin{deluxetable}{rrrrccccc}
\tablenum{1}
\tablewidth{0pc}
\tablecaption{Log of the Keck/HIRES Observations of M5 Stars}
\tablehead{
\colhead{Star} & \colhead{V} & \colhead{$\sigma$}
& \colhead{B-V} & {$\sigma$} & \colhead{(Date-UT)} & \colhead{Exp(min)} &
\colhead{Total(min)}  &\colhead{S/N} 
} 
\startdata 
%Star 	V       sig	B-V     sig	  Date          exp    Total      S/N
5287 &18.0224 &0.0036 &0.5108 &0.0074 & 1999 May 10 & 5$\times$45 &     &  \\
     &        &       &       &       & 1999 May 11 & 1$\times$45 & 270 & 39 \\
5303 &18.0243 &0.0043 &0.5145 &0.0074 & 1999 May 10 & 3$\times$50 &     & \\
     &        &       &       &       &             & 1$\times$45 & 195 & 38 \\
5323 &18.0270 &0.0036 &0.5187 &0.0073 & 1999 May 11 & 6$\times$45 & 270 & 37 \\
5351 &18.0319 &0.0036 &0.5240 &0.0080 & 1999 Jun 08 & 1$\times$45 &     & \\
     &	      &       &       &       &             & 5$\times$60 & 345 & 35 \\
5364 &18.0338 &0.0036 &0.5128 &0.0058 & 1999 Jun 07 & 2$\times$45 &     & \\
     &        &       &       &       &             & 4$\times$60 & 330 & 32 \\
\enddata
\end{deluxetable}
%\end{document}

\clearpage

%\documentclass[12pt,preprint]{aastex}
%\begin{document}
\begin{deluxetable}{lrc}
\tablenum{2}
\tablewidth{0pc}
\tablecaption{Stellar Parameters}
\tablehead{
\colhead{Star} & \colhead{$T_{\rm eff}(K)$} & \colhead{log g} 
}
\startdata
%star     REF. Teff   log g 
M5:5287  & C83  5906 & 3.92  \\
         & K93  6039 & 4.00  \\
	 & RM05 5830 & 3.84  \\
M5:5303  & C83  5890 & 3.91  \\
         & K93  6024 & 3.95  \\
         & RM05 5815 & 3.87  \\
M5:5323  & C83  5871 & 3.90  \\
         & K93  6007 & 3.98  \\
         & RM05 5798 & 3.82  \\
M5:5351  & C83  5848 & 3.89  \\
         & K93  5986 & 3.97  \\
         & RM05 5775 & 3.81  \\
M5:5364  & C83  5897 & 3.92  \\
         & K93  6031 & 4.00  \\
         & RM05 5822 & 3.84  \\
\enddata
\end{deluxetable}
%\end{document}

\clearpage

%\documentclass[12pt,preprint]{aastex}
%\begin{document}
\begin{deluxetable}{lrrrccccccc}
\tablenum{3}
\tablewidth{0pc}
\tablecaption{Lithium Abundances}
\tablehead{
\colhead{Star} & \colhead{T[K]} & \colhead{T[C]}
& \colhead{T[R]} 
& \colhead{A(Li)$_K$} & \colhead{A(Li)$_C$} & \colhead{A(Li)$_R$} 
& \colhead{EQW}
& \colhead{A(Li)$_K$} & \colhead{A(Li)$_C$} & \colhead{A(Li)$_R$} 
}
\startdata
%STAR	TK	TC	TRM   ABK    ABC    ABR   EQW  AB-K	AB-C	AB-R
5287 & 6039 & 5906 & 5830 & 2.53 & 2.42 & 2.37 & 53.3 & 2.57 & 2.47 & 2.41 \\
5303 & 6024 & 5890 & 5815 & 2.55 & 2.45 & 2.45 & 47.9 & 2.51 & 2.40 & 2.33 \\
5323 & 6007 & 5871 & 5798 & 2.48 & 2.35 & 2.35 & 53.7 & 2.55 & 2.45 & 2.38 \\
5351 & 5986 & 5848 & 5775 & 2.30 & 2.17 & 2.10 & 33.8 & 2.30 & 2.18 & 2.13 \\
5364 & 6031 & 5897 & 5822 & 2.38 & 2.28 & 2.24 & 32.8 & 2.32 & 2.21 & 2.14 \\
\hline
RANGE & 53   &  58  &   55 & 0.25 & 0.28 & 0.35 & 20.9 & 0.27 & 0.29 & 0.28 \\
\enddata
\end{deluxetable}
%\end{document}

\clearpage

%\documentclass[12pt,preprint]{aastex}
%\begin{document}
\begin{deluxetable}{llcc}
\tablenum{4}
\tablewidth{0pc}
\tablecaption{Globular Cluster Ages and Metallicities}
\tablehead{
\colhead{NGC} & \colhead{Name} & \colhead{Age (Gyr)} & \colhead{[Fe/H]} 
}
\startdata
%Cluster 	name	Age	[Fe/H]
NGC 104	 &	47 Tuc & 13.54$\pm$0.90& $-$0.747$\pm$0.003 \\ 
NGC 5904 &	M 5    & 12.75$\pm$0.80& $-$1.259$\pm$0.003 \\
NGC 6121 &	M 4    & 13.01$\pm$1.01& $-$1.166$\pm$0.004 \\
NGC 6341 &	M 92   & 13.30$\pm$0.60& $-$2.239$\pm$0.028 \\
NGC 6397 &	...    & 14.21$\pm$0.69& $-$1.994$\pm$0.004 \\
NGC 6752 &	...    & 13.47$\pm$0.67& $-$1.583$\pm$0.003 \\
\enddata
\end{deluxetable}
%\end{document}

\end{document}